\documentclass [preprint,aps,prb,showpacs,amsmath,amssymb]{revtex4}
\usepackage{graphicx}
\usepackage{dcolumn}
\usepackage{bm}

\begin{document} 

\title{Second-harmonic generation and linear electro-optical coefficients of BN nanotubes}  
\author{G.Y. Guo\footnote{Electronic address: 
gyguo@phys.ntu.edu.tw} and J.C. Lin}
\address{Department of Physics, National Taiwan University, Taipei, Taiwan 106, 
Republic of China}
\date{\today}

\begin{abstract}
A systematic {\it ab initio} study of the second-order nonlinear optical  
properties of BN nanotubes within density functional 
theory in the local density approximation has been performed.
Highly accurate full-potential projector augmented-wave method was used.
Specifically, the second-harmonic generation ($\chi_{abc}^{(2)}$) and linear electro-optical 
($r_{abc}$) coefficients of a large number of the single-walled zigzag, armchair 
and chiral BN nanotubes (BN-NT) as well as the double-walled
zigzag (12,0)@(20,0) BN nanotube and the single-walled zigzag (12,0) BN-NT bundle 
have been calculated. 
Importantly, unlike carbon nanotubes, both the zigzag and chiral BN-NTs are found to
exhibit large second-order nonlinear optical behavior with the $\chi_{abc}^{(2)}$
and $r_{abc}$ coefficients being up to thirty times larger than that of 
bulk BN in both zinc-blende and wurtzite structures, indicating that
BN-NTs are promising materials for nonlinear optical and opto-electric applications. 
Though the interwall interaction in the double-walled BN-NTs is found to reduce
the second-order nonlinear optical coefficients significantly, the interwall interaction
in the single-walled BN-NT bundle has essentially no effect on the nonlinear optical 
properties. The prominant features in the spectra of $\chi_{abc}^{(2)}(-2\omega,\omega,\omega)$ 
of the BN-NTs are successfully correlated with the features in the 
linear optical dielectric function $\varepsilon (\omega)$ in terms of single-photon
and two-photon resonances.

\end{abstract}


\maketitle

\section{Introduction}

Since their discovery in 1991~\cite{iij91}, carbon nanotubes (CNTs) have attracted 
considerable interest worldwide because of their unusual properties and great 
potentials for technological applications. For example, because of their 
one-dimensional character, metallic CNTs are quantum wires 
that may exhibit exotic Luttinger-liquid behavior rather than usual
Fermi-liquid behavior in normal metal wires.~\cite{boc99} 
It was also predicted that nanotori formed from metallic CNTs may 
exhibit giant paramagnetic moments.~\cite{liu02a} 
Furthermore, chiral CNTs 
are expected to exhibit a number of unusual optical properties such as 
optical activity, circular dichroism and second harmonic 
generation (see ~\cite{guo04a,guo04b} and references therein).

Soon after the discovery of CNTs it became obvious that similar nanostructures
could be formed by other elements and compounds which form layered structures
bearing some resemblance to graphite. For example, hexagonal BN ($h$-BN) was
predicted on the basis of theoretical calculations~\cite{rub94,bla94}
to be capable of forming nanotubes, a prediction which was later confirmed
experimentally by the synthesis of such nanotubes.~\cite{cho95} 
Both single-walled and multiwalled BN nanotubes (BN-NT) can now be readily 
synthesized.~\cite{lee01} Though CNTs continue to attract great interest,
other nanotubes such as BN-NTs are interesting in their own right and may 
be able to offer different possibilities for technological applications
that CNTs cannot provide. In particular, as far as the optical and 
opto-electronic applications of nanotubes are concerned, BN-NT could be
superior to CNTs because BN-NTs are uniformly insulating, independent of
their chirality. 
Furthermore, BN-NTs tend to have a zigzag structure.~\cite{lee01}
Though it is interesting that, depending on their chirality, 
CNTs can be metallic or semiconducting or insulating~\cite{guo04a}, it is still impossible
to grow CNTs with a pre-specified chirality at present.
Finally, recent experiments indicate that BN-NTs exhibit stronger resistance
to oxidation at high temperatures than CNTs~\cite{che04}
 
Therefore, the electronic, optical and other properties of both
single-walled and multiwalled BN-NTs are interesting and have
been intensively studied theoretically in recent years (see, e.g.,
~\cite{rub94,bla94,alo01,kim01,mel02,oka02,che02,akd03,xia03,ng04}). 
In particular, Chen {\it et al.}~\cite{che02} calculated the transverse dielectric
function of bundles of single-walled BN-NTs using a tight-binding model.
Ng and Zhang~\cite{ng04} calculated the optical absorption spectra of 
single-walled BN-NTs within a time-dependent localized-density-matrix 
approach based on a semiempirical Hamiltonian.
Despite of these intensive theoretical studies, only few {\it ab initio} 
calculations of the optical properties of BN-NTs have been reported~\cite{mar04} 
because of the heavy demand of the computing 
resources. Semiempirical tight-binding model is known to describe well only
the electronic excitations near the band gap of the large radius BN-NTs.
Systematic {\it ab initio} calculations of the optical
properties are thus needed in order to quantitatively interpret the optical
experiments and to predict the BN-NTs with desired optical properties. 
Therefore, we have recently carried out a series of {\it ab initio} calculations~\cite{guo05}
in order to analysize the band structure and linear optical features of all the three types of
the BN-NTs and their possible dependence on diameter and chirality.
In this work, we investigate the second-order optical susceptibility and also
linear electro-optical coefficient of the BN-NTs. The primary objective is to find out 
the features and magnitude of the second-harmonic generation and linear electro-optical
coefficients of the BN-NTs in order to see whether they have any potential
applications in nonlinear optical and electro-optical devices such as second-harmonic
generation, sum-frequency generation and electrical optical switch. 
The second objective is to identify characteristic differences in 
nonlinear optical properties between BN-NTs and CNTs.

The rest of this paper is organized as follows. In Sec. II, the theoretical
approach and computational details are briefly described. 
In Sec. III, the calculated second-order nonlinear optical susceptibility and linear
electro-optical coefficients of $h$-BN, single
hexagonal BN sheet, and BN nanotubes are presented and analysized. 
Finally, in Sec. IV, a summary is given.

\section{Theory and computational method}

BN-NTs can be considered as
a layer of graphitic BN sheet rolled up into a cylinder, and the structure of
a BN-NT is completely specified by  
the chiral vector which is given in term of a pair of integers 
($n$,$m$).~\cite{rub94,bar99}  
As for CNTs, BN-NTs are classified into three types, namely, armchair ($n$,$n$) 
nanotubes, zigzag ($n$,0) nanotubes, and chiral ($n$,$m$) nanotubes with 
$n \neq m$.~\cite{rub94,bar99}
We consider a large number of representative BN-NTs with a range of
diameters from all three types, namely, the zigzag
[($n$,0), $n$ = 5, 6, 8, 9, 12, 13, 15, 16, 17, 20, 21, 24, 25, 27], armchair 
[($n$,$n$), $n$ = 4, 5, 6, 8, 12, 15], 
and chiral (4,2), (6,4), (8,4), (10,5) BN-NTs. The double-walled
zigzag (12,0)@(20,0) BN nanotube and the single-walled zigzag (12,0) BN-NT
bundle have also been investigated to see 
the effects of interwall interaction. 
Our {\it ab initio} calculations for the BN-NTs were performed using
highly accurate full-potential projector augmented-wave (PAW) 
method~\cite{blo94}, as implemented in the VASP package~\cite{kre93}.
They are based on density functional theory (DFT) with the local density approximation (LDA). 
A supercell geometry was adopted so that the nanotubes are aligned in a
square array with the closest distance between adjacent nanotubes being
at least 6 \AA. 
A large plane-wave cut-off of 450 eV was used throughout.

Firstly, the ideal nanotubes were constructed by rolling-up a 
hexagonal BN sheet. 
Their atomic positions and lattice constants were then fully relaxed by a conjugate 
gradient technique. Theoretical equilibrium nanotube structures were obtained
when the forces acting on all the atoms and the uniaxial stress were less than 0.03 eV/\AA$ $
and 2.0 kBar, respectively. 
The theoretical equilibrium lattice constants $T$
and curvature energies $E_c$ (total energy relative to that of
single BN sheet) as well as the computational details have been reported before~\cite{guo05}.

The self-consistent electronic band structure calculations were then
carried out for the theoretically determined BN-NT structures. 
In this work, the non-linear optical properties were calculated based on the
independent-particle approximation, i.e., the excitonic effects and the 
local-field corrections were neglected. As reported before~\cite{guo05},
the dielectric function
of $h$-BN calculated within the single-electron picture 
are in reasonably good
agreement with experiments. Therefore, it might be expected that the 
independent-particle approximation could work rather well for the non-linear optical
properties of the BN-NTs too.
Following previous nonlinear optical calculations~\cite{dua99,guo04a}, the imaginary
part of the second-order
optical susceptibility due to direction interband transitons is given by~\cite{gha90}
\begin{equation}
 \chi^{''(2)}_{abc}(-2\omega,\omega,\omega) = \chi^{''(2)}_{abc,VE}(-2\omega,\omega,\omega)
 + \chi^{''(2)}_{abc,VH}(-2\omega,\omega,\omega) 
\end{equation}
where the contribution due to the so-called virtual-electron (VE) process is
\begin{eqnarray} \label {eq:VE}
\chi^{''(2)}_{abc,VE} = -\frac{\pi}{2\Omega}\sum_{i\in VB} \sum_{j,l\in CB}
 \sum_{\bf k} w_{\bf k}
 \{ \frac{Im[p_{jl}^a\langle p_{li}^bp_{ij}^c\rangle]}{\epsilon^3_{li}(\epsilon_{li}+
 \epsilon_{ji})}\delta(\epsilon_{li}-\omega) \nonumber \\
 -\frac{Im[p_{ij}^a\langle p_{jl}^bp_{li}^c\rangle]}{\epsilon^3_{li}(2\epsilon_{li}-
  \epsilon_{ji})}\delta(\epsilon_{li}-\omega) 
  +\frac{16Im[p_{ij}^a\langle p_{jl}^bp_{li}^c\rangle]}{\epsilon^3_{ji}(2\epsilon^3_{li}-
  \epsilon^3_{ji})}\delta(\epsilon_{ji}-2\omega) \}
\end{eqnarray}
and that due to the virtual-hole (VH) process
\begin{eqnarray} \label {eq:VH}
\chi^{''(2)}_{abc,VH} = \frac{\pi}{2\Omega}\sum_{i,l\in VB} \sum_{j\in CB}
 \sum_{\bf k} w_{\bf k}
 \{ \frac{Im[p_{li}^a\langle p_{ij}^bp_{jl}^c\rangle]}{\epsilon^3_{jl}(\epsilon_{jl}+
 \epsilon_{ji})}\delta(\epsilon_{jl}-\omega) \nonumber \\
 -\frac{Im[p_{ij}^a\langle p_{jl}^bp_{li}^c\rangle]}{\epsilon^3_{jl}(2\epsilon_{jl}-
 \epsilon^3_{ji})}\delta(\epsilon_{jl}-\omega) 
  +\frac{16Im[p_{ij}^a \langle p_{jl}^bp_{li}^c\rangle]}{\epsilon^3_{ji}(2\epsilon_{jl}-
 \epsilon_{ji})}\delta(\epsilon_{ji}-2\omega) \}.
\end{eqnarray}
Here $\epsilon_{ji} = \epsilon_{{\bf k}j}-\epsilon_{{\bf k}i}$ and
$\langle p_{jl}^bp_{li}^c\rangle = \frac{1}{2}(p_{jl}^bp_{li}^c+p_{li}^bp_{jl}^c)$.
The dipole transition matrix elements $p_{ij}^a = <{\bf k}j|\hat{p}_a|{\bf k}i>$ were obtained from
the self-consistent band structures within the PAW formalism ~\cite{ado01}.
The real part of the second-order optical susceptibility is then obtained from
$\chi''^{(2)}_{abc}$ by
a Kramer-Kronig transformation
\begin{equation}
\chi'^{(2)}(-2\omega,\omega,\omega) = \frac{2}{\pi}{\bf P} \int_0^{\infty}d\omega'
 \frac{\omega'\chi''^{(2)}(2\omega',\omega',\omega')}{\omega'^2-\omega^2}.
\end{equation}

The linear electro-optic coefficient $r_{abc}(\omega)$ is connected to the 
second-order optical susceptibility $\chi^{(2)}_{abc}(-\omega,\omega,0)$ through 
the relation~\cite{hug96}
\begin{equation}
\chi^{(2)}_{abc}(-\omega,\omega,0) = -\frac{1}{2} n^2_{a}(\omega)n^2_{b}(\omega)r_{abc}(\omega)
\end{equation}
where $n(\omega)$ is the refraction index in the $a$-direction. In the zero frequency limit,
\begin{equation}
\lim_{\omega\rightarrow 0}\chi^{(2)}_{abc}(-2\omega,\omega,\omega) 
=\lim_{\omega\rightarrow 0}\chi^{(2)}_{abc}(-\omega,\omega,0).
\end{equation}
Therefore, 
\begin{equation}
r_{abc}(0) = -\frac{2}{n^2_{a}(0)n^2_{b}(0)}\lim_{\omega\rightarrow 0}\chi^{(2)}_{abc}(-2\omega,\omega,\omega).
\end{equation}
Furthermore, for the photon energy $\hbar\omega$ well below the band gap,
the linear electro-optic coefficient $r_{abc}(\omega) \approx r_{abc}(0)$
because $\chi^{(2)}_{abc}(-2\omega,\omega,\omega)$ and $n(\omega)$ 
are nearly constant in this low frequency region, as shown in the next Sec. and 
in Ref. ~\cite{guo05}.
 
In the present calculations, the $\delta$-function in Equs. 2-3
is approximated by a Gaussian function with $\Gamma = 0.2$ eV.
The same $k$-point grid as in the DOS calculation is used. 
Furthermore, to ensure that $\varepsilon'$ calculated via Kramer-Kronig
transformation (Equ. 4) is reliable, at least ten energy bands 
per atom are included in the present optical calculations. The unit cell
volume $\Omega$ in Equs. 2-3 is not well defined for nanotubes.
Therefore, like the previous calculations~\cite{guo04a,guo04b,guo05}, we used the
effective unit cell volume of the nanotubes rather than the volume of the supercells
which is arbitrary. The effective unit cell of a nanotube is given by
$\Omega = \pi[(D/2+d/2)^2-(D/2-d/2)^2]T = \pi DdT$ where $d$ is the thickness of the
nanotube cylinder which is set to the interlayer distance of $h$-BN 
(3.28 \AA~\cite{guo05}). $D$ and $T$ are the diameter and length of 
translational vector of the nanotube~\cite{guo05}, respectively.  

\section{Results and discussion}

\subsection{Hexagonal BN and single BN sheet}

For comparison with the BN-NTs, we first investigated
the second-order nonlinear optical susceptibility of 
both $h$-BN and an isolated honeycomb BN sheet. 
The isolated BN sheet is simulated
by a slab-supercell approach with an inter-sheet distance of 6.5 \AA.
The theoretically determined lattice constants ($a = 2.486$ \AA $ $ and
$c = 6.562$ \AA $ $ for $h$-BN and $a = 2.485$ \AA $ $ 
for the BN sheet)~\cite{guo05} 
were used. Note that the theoretical lattice
constants of $h$-BN agree rather well (within 1.5 \%) with the 
experimental values ($a = 2.50$ \AA $ $ and $c = 6.65$ \AA)~\cite{cap96}.
Interestingly, we find numerically that although $h$-BN has zero 
second-order nonlinear optical susceptibility, 
the $\chi^{(2)}_{aab}$, $\chi^{(2)}_{baa}$ and 
$\chi^{(2)}_{bbb}$ for the isolated BN sheet are nonzero. Here
$a$ and $b$ denote the two Cartesian coordinates within the BN layer.
Moreoever, $\chi^{(2)}_{baa} = \chi^{(2)}_{aab} $ 
and $\chi^{(2)}_{bbb} = -\chi^{(2)}_{aab} $. This is consistent
with the symmetry consideration, demonstrating that our 
numerical method and calculations are qualitatively correct.
Bulk $h$-BN should have zero second-order nonlinear optical 
susceptibility due to its spatial inversion symmetry. On the other hand,
the isolated BN sheet does not have the spatial inversion symmetry ($D_{6h}^4$)
and its symmetry class is $D_{3h}$ ($P\bar{6}m2$). Therefore, the isolated
BN sheet has nonzero $\chi^{(2)}_{aab}$, $\chi^{(2)}_{baa}$ and
$\chi^{(2)}_{bbb}$ from the symmetry consideration.~\cite{boy03} 

\begin{figure}
\includegraphics[width=8cm]{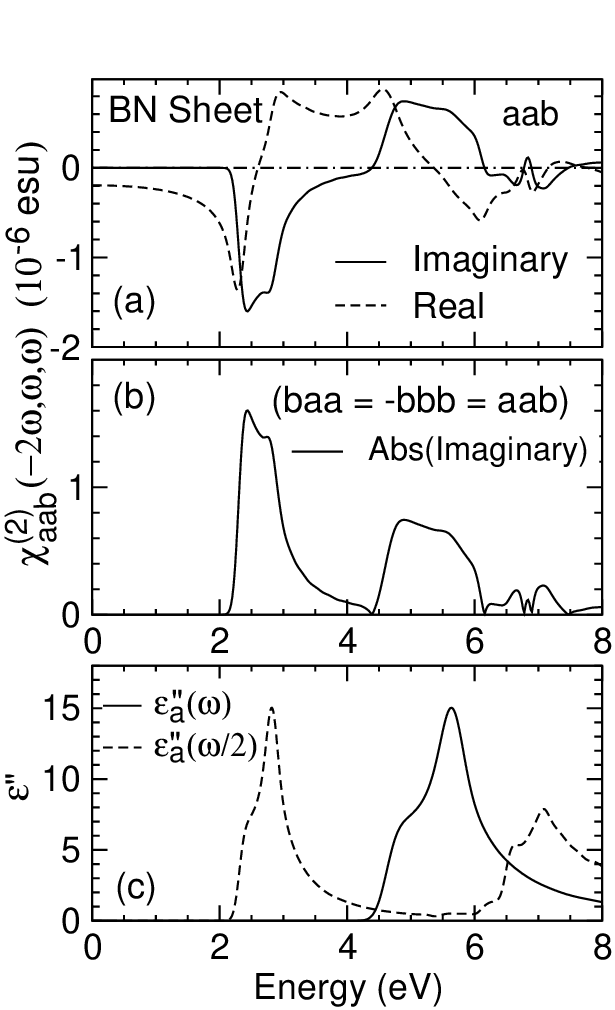}
\caption{\label{fig1} (a) Real and imaginary parts as well as (b)
the absolute value of the imaginary part of $\chi^{(2)}_{aab}(-2\omega,\omega,\omega)$
of the isolated BN sheet. In (c), $\varepsilon_a''(\omega)$
and $\varepsilon_a''(\omega/2)$ (imaginary part of the
dielectric function) from Ref. ~\cite{guo05} are plotted.}
\end{figure}

In Fig. 1, the calculated real and imaginary parts as well as
the absolute value of the imaginary part of $\chi^{(2)}_{aab}(-2\omega,\omega,\omega)$
of the single BN sheet are shown. 
In the calculations, a $k$-point grid of $100\times 100\times 2$ for
the honeycomb BN sheet is used. It is clear from Fig. 1 that the
second harmonic generation (SHG) coefficient 
$\chi^{(2)}_{aab}(-2\omega,\omega,\omega)$ is significant in the entire
range of the optical photon energy ($\hbar\omega$).
Furthermore, for the photon energy smaller than 2.2 eV, the $\chi^{(2)}_{aab}$
is purely dispersive (i.e., real and lossless) (Fig. 1a), suggesting that
the BN sheet has potential application in nonlinear optical devices. 
Note that since the BN sheet has a theoretical band gap ($E_g$) 
of $\sim$4.5 eV, the absorptive (imaginary) part of the $\chi^{(2)}_{aab}$ 
becomes nonzero only above the half of the band gap (i.e., $\sim$2.2 eV). 
The real part of the $\chi^{(2)}_{aab}$ remains nearly constant at 
low photon energies up to 1.0 eV, then increases steadily 
in magnitude as the photon energy increases, and finally peaks at the
absorption edge of $\sim$2.2 eV (Fig. 1a). In the energy range from 2.5 to 
5.5 eV, the real part of the $\chi^{(2)}_{aab}$ becomes positive and
forms a broad double peak structure. Beyond 5.5 eV, it becomes negative
again and its magnitude gradually deminishes as the photon energy
further increases (Fig. 1a). 

The absorptive part of the $\chi^{(2)}_{aab}$ 
is nonzero only above $\sim$2.2 eV, and looks like a Lorentzian 
oscillation between 2.2 and 6.0 eV with one sharp negative peak at $\sim$ 
2.5 eV and one broad positive peak around 5.5 eV (Fig. 1a). 
It is clear from Equs. 2-3 that the
calculated $\chi^{(2)}_{aab}$ spectra can have pronounced features due
to both single- and double-frequency resonant terms. To
analyze the features in the calculated $\chi^{(2)}$ spectra, it is helpful
to compare the absolute value of $\chi''^{(2)}$ (Fig 1b) with the 
absorptive part of the corresponding dielectric function
$\varepsilon''$.~\cite{sip98} Therefore, the calculated $\varepsilon''$
from our previous publication~\cite{guo05} are shown in Fig. 1c as a function
of both $\omega /2$ and $\omega$. Clearly, the first sharp peak at $\sim$2.5 eV
is due to two-photon resonances [cf. $\varepsilon_a''(\omega/2)$] whilst in contrast,
the second broad peak around 5.5 eV comes from the single-photon resonances
[cf. $\varepsilon_a''(\omega)$]. Nevertheless, both single- and double-photon resonances
involve only interband $\pi \rightarrow \pi^{\ast}$
and $\sigma \rightarrow \sigma^{\ast}$ optical transitions for the
electric field vector ${\bf E}$ polarized parallel to the BN layer
($E \parallel \hat{a}$)~\cite{guo05}.    

\begin{table}
\caption{Calculated static refraction index $n$, second-order optical
susceptibility $\chi^{(2)}(0)$ and linear electro-optical
coefficient $r_{abc}$ of the isolated BN sheet.
}
\begin{ruledtabular}

\begin{tabular}{ c c c}
      $n_a$ ($n_c$) & $\chi^{(2)}_{bbb}$, $\chi^{(2)}_{aab}$ (pm/V)
      & $r_{bbb}$, $r_{aab}$ (pm/V) \\ \hline
 2.19 (1.63) & 81.3, -81.3 & -3.56, 3.56        \\ 
\end{tabular}
\end{ruledtabular}
\end{table}

In Table I, the calculated zero frequency linear electro-optic coefficient
$r(0)$ as well as the corresponding second-order nonlinear optical 
susceptibility $\chi^{(2)}(0,0,0)$ and the static refraction
index $n(0)$ are listed. The refraction index $n(0) (=\sqrt{\varepsilon(0)})$
is derived from the calculated static dielectric constant
$\varepsilon(0)$ which has been reported in our recent publication~\cite{guo05}.
Note that the $r(0)$ and $\chi^{(2)}(0,0,0)$ are listed in the SI
pm/V unit, and 1 pm/V = $4.1888\times10^{-8}$ esu.
Significantly, the static second-order optical susceptibility for the isolated
BN sheet is nearly thirty times larger than that of BN in both the zinc-blende and 
wurtzite structures~\cite{gav00}. 

\subsection{Second-order optical susceptibility}

We have explicitly calculated the second-order optical susceptibility for the
zigzag (5,0), (6,0), (8,0), (9,0), (12,0), (13,0), (15,0), (16,0), (17,0), (20,0), (21,0),
(24,0), (25,0), (27,0), armchair (3,3), (4,4), (5,5), (6,6), (8,8), (12,12), (15,15), 
and chiral (4,2), (6,4), (8,4), (10,5) BN-NTs. 
In the calculations, a uniform grid ($1\times 1\times m$)
along the tube axis (z-axis) was used. The number $m$ is 200, 120, 150, 60, 100, 100, 80,
120, 80, 80, 60, 80, 60, 60 for the sigzag
BN-NTs, respectively, and 60, 50, 40, 60 for the chiral BN-NTs, respectively,
In the case of CNTs, only the chiral nanotubes would
exhibit second-order nonlinear optical behavior with
two nonvanishing components of $xyz$ and $xzy$ of $\chi^{(2)}$.~\cite{guo04a}
Here $z$ refers to the coordinate along the tube axis whilst
$x$ and $y$ denote the two coordinates that are perpendicular to the
tube axis. As for CNTs, the armchair BN-NTs are found not to have any nonzero components
of $\chi^{(2)}$. 
In contrast, both the zigzag and chiral BN-NTs are found to 
show second-order nonlinear optical behavior.
Specifically, all the zigzag BN-NTs except (5,0), (9,0) and (27,0), have six nonvanishing
components of the second-order optical susceptibility, namely,
$xzx$, $xxz$, $yyz$, $zxx$, $zyy$, $zzz$. Nevertheless, these components are not completely 
independent of each other. In particular,
$\chi^{(2)}_{xxz}= \chi^{(2)}_{yyz} = \chi^{(2)}_{xzx}$, and 
$\chi^{(2)}_{zyy} = \chi^{(2)}_{zxx}$.
This finding is especially important for the application of BN-NTs in
nonlinear optical devices because most BN-NTs tend to have a zigzag structure,~\cite{lee01} 
as mentioned before.
The chiral BN-NTs have eight nonvanishing
components of the second-order optical susceptibility with two additional nonzero
components being $xyz$ and $yzx$. Note that $\chi^{(2)}_{yzx} = -\chi^{(2)}_{xyz}$.
These numerical findings are consistent with the consideration of the symmetry of
the BN nanotubes. The point symmetry groups of the BN-NTs~\cite{alo01} are 
$C_{2nv}$ for zigzag ($n$,0) nanotubes, $C_{2nh}$ for armchair ($n,n$) nanotubes, and
$C_N$ for chiral ($n,m$) nanotubes where $N$ = $2(n^2+m^2+nm)/d_R$ with $d_R$ being
the greatest common divisor of $2n+m$ and $2m+n$. Therefore, these symmetries would
dictate~\cite{boy03} that all the components vanish for armchair ($n,n$) nanotubes, and that
nonvanishing components for zigzag ($n$,0) nanotubes are $xzx$ = $yzy$, $xxz$ = $yyz$,
$zxx = zyy$, $zzz$, as well as that nonvanishing components of chiral ($n,m$) nanotubes 
include all that for zigzag ($n$,0) nanotubes plus $xyz$ = -$yzx$.  
We don't know at the moment why the (5,0), (9,0) and (27,0) BN-NTs are 
found numerically to
have no nonzero components though they don't have to from their symmetry point of view.

\begin{figure}[tb]
\includegraphics[width=16cm]{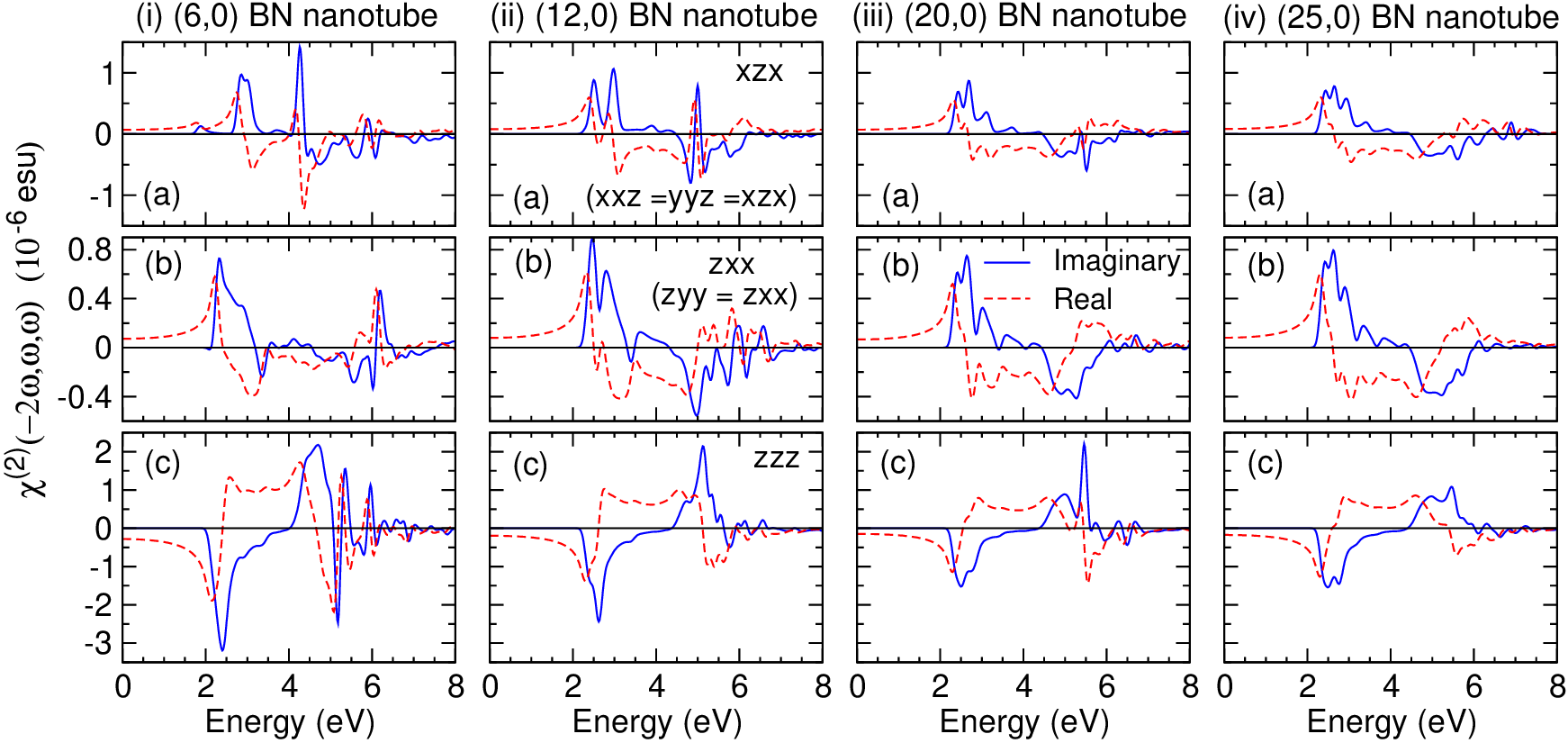}
\caption{\label{fig2} (Color online) Real and imaginary parts
of $\chi^{(2)}(-2\omega,\omega,\omega)$ of the zigzag (6,0), (12,0),
(20,0) and (25,0) BN nanotubes.}
\end{figure}

In Fig. 2, the calculated real and imaginary parts of the second-order
optical susceptibility $\chi^{(2)}(-2\omega,\omega,\omega)$ for the
four selected zigzag nanotubes [(6,0), (12,0), (20,0), and (25,0)] are shown.
As for the single BN sheet, the second harmonic generation coefficients
$\chi^{(2)}(-2\omega,\omega,\omega)$ for the zigzag BN-NTs are significant in the entire
range of the optical photon energy ($\hbar\omega$). Indeed, they are generally
more than ten times larger than that of bulk BN in the zinc-blende and wurtzite 
structures.~\cite{gav00} 
Moreover, for the photon energy smaller than 2.0 eV, the $\chi^{(2)}$
is purely dispersive (i.e., real and lossless) (Fig. 2). 
Interestingly, the magnitude of $\chi^{(2)}_{zzz}$ is the largest and in
general about three times larger than that of the other nonvanishing components (Fig. 2).
For $\chi^{(2)}_{zzz}$, the electric field of both the incoming and outgoing
photons is polarized parallel to the tube axis and thus the electric dipolarization
effect would be essentially zero~\cite{guo05}. This may be particularly important
for nonlinear optical applications. 
We also note that the shape of the spectra of $\chi^{(2)}_{zzz}$ and $\chi^{(2)}_{zxx}$
for all the zigzag BN-NTs look very similar, except the difference in sign.
For the zigzag BN-NTs with a larger diameter such as (12,0), (20,0) and (25,0), 
the spectrum of $\chi^{(2)}_{xzx}$ is also similar to that of $\chi^{(2)}_{zxx}$ (Fig. 2).
The magnitude of all the components decreases somewhat as the diameter of the tubes increases
[e.g., from (6,0) to (12,0)], but however, becomes stable as the diameter further increases 
[see, e.g., (20,0) and (25,0) in Fig. 2]. Furthermore, both the magnitude and shape
of the $\chi^{(2)}_{zzz}$ spectrum for the zigzag BN-NTs with a larger
diameter, e.g., (25,0), approach to that of the single BN sheet (Fig. 1 and Fig. 2), as they should. 

\begin{figure}[tb]
\includegraphics[width=8cm]{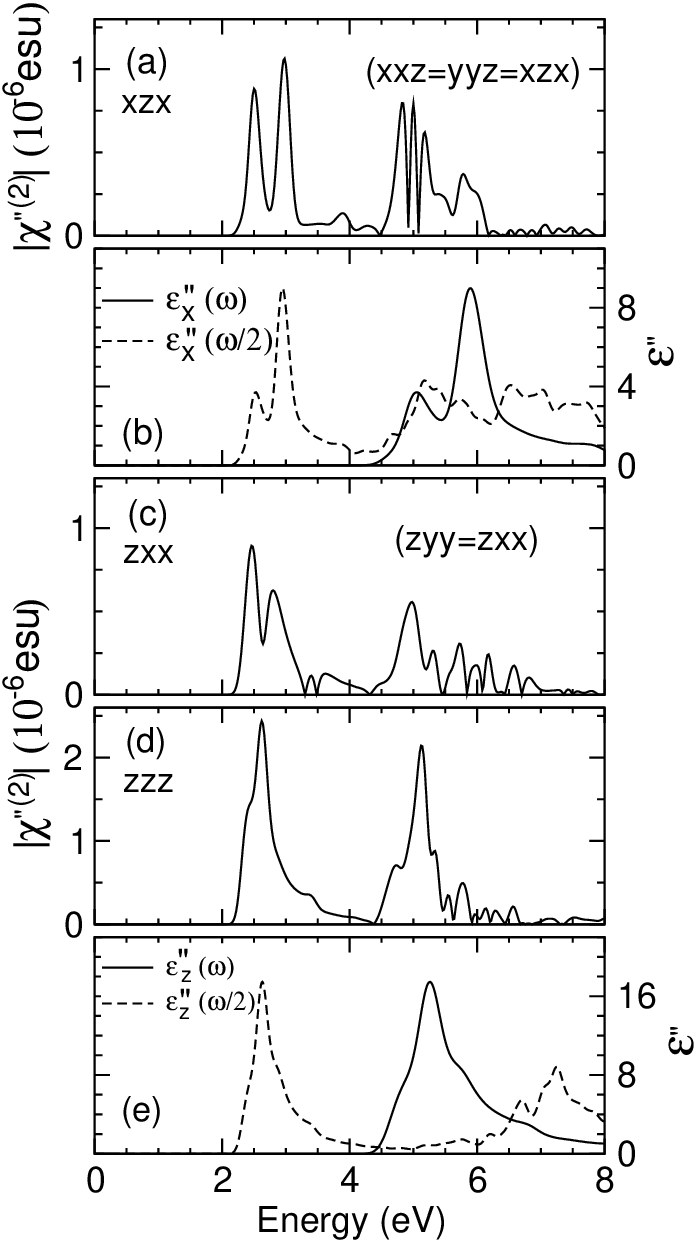}
\caption{\label{fig3} Absolute value of the imaginary part 
of $\chi^{(2)}(-2\omega,\omega,\omega)$ (a, c, d) as well as
$\varepsilon_a''(\omega)$ and $\varepsilon_a''(\omega/2)$ (imaginary part
of the dielectric function) (b, e) from Ref. ~\cite{guo05} of the zigzag (12,0)
BN nanotube.}
\end{figure}

As for the single BN sheet, in order to understand the features in the 
calculated $\chi^{(2)}$ spectra of the zigzag BN-NTs, the absolute values of the imaginary part
$\chi''^{(2)}$ of all the nonzero components of the (12,0) BN-NT are plotted and compared with the
absorptive part of the dielectric function
$\varepsilon''$ from our previous publication~\cite{guo05} in Fig. 3.  
Strikingly, the first prominent peak between 2.0 and 4.0 eV 
in the $\chi_{zzz}''^{(2)}$ spectrum is almost identical to the first peak in
the $\varepsilon_z''(\omega/2)$ (see Fig. 3d-e), indicating that it
is due to two-photon resonances. In contrast, the second peak between 4.5 and 5.5 eV
in the $\chi_{zzz}''^{(2)}$ spectrum is very similar to the
first peak in the $\varepsilon_z''(\omega)$, suggesting that 
it is caused by the single-photon resonances. Nevertheless, 
both these single- and double-photon resonances
involve only interband $\pi \rightarrow \pi^{\ast}$
and $\sigma \rightarrow \sigma^{\ast}$ optical transitions for the
electric field vector ${\bf E}$ polarized parallel to the tube axis 
($E \parallel \hat{z}$)~\cite{guo05}.  
Fig. 3 also shows that the double-peak structure between 2.0 and 4.0 eV in
both the $\chi_{xzx}''^{(2)}$ and $\chi_{zxx}''^{(2)}$ spectra is mainly due to the
two-photon resonances with $E\perp \hat{z}$ [cf. $\varepsilon_x''(\omega/2)$]
(see Fig. 3a-c), while, in contrast, the second feature in the photon energies above
4.5 eV perhaps comes predominantly from the the single-photon resonances
for $E\perp \hat{z}$ [cf. $\varepsilon_x''(\omega)$]. This conclusion is further 
supported by the fact that the magnitude of $\varepsilon_x''(\omega)$ ($\varepsilon_x''(\omega/2)$)
is only about half of that of $\varepsilon_z''(\omega)$ ($\varepsilon_z''(\omega/2)$),
and concurrently the magnitude of $\chi_{xzx}''^{(2)}$ and $\chi_{zxx}''^{(2)}$ 
is only about half of that of $\chi_{zzz}''^{(2)}$ too. 

\begin{figure}[tb]
\includegraphics[width=16cm]{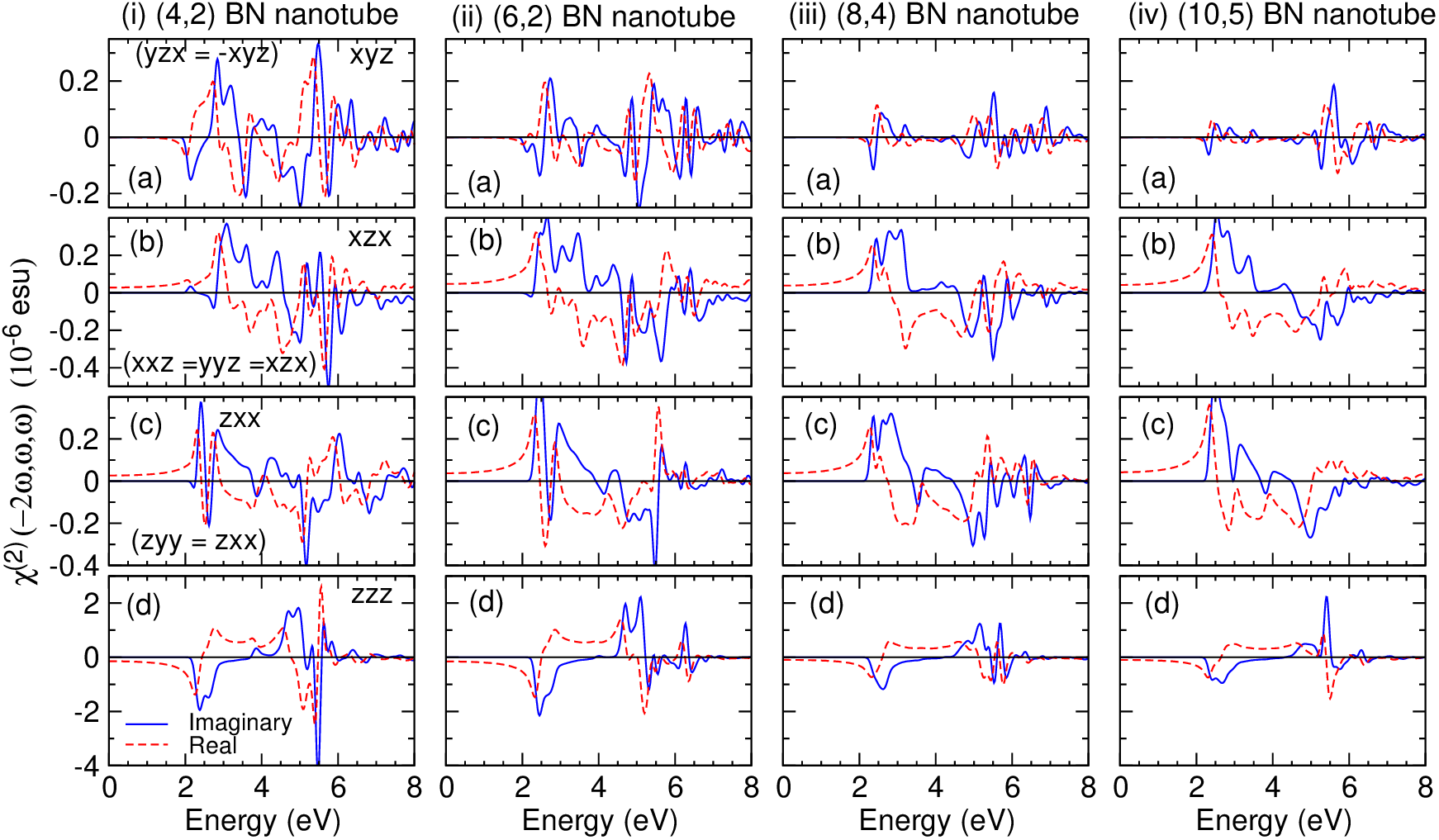}
\caption{\label{fig4} (Color online) Real and imaginary parts
of $\chi^{(2)}(-2\omega,\omega,\omega)$ of the chiral (4,2), (6,2),
(8,4) and (10,5) BN nanotubes.}
\end{figure}

In Fig. 4, the calculated real and imaginary parts of the second-order
optical susceptibility $\chi^{(2)}(-2\omega,\omega,\omega)$ for all
the four chiral nanotubes [(4,2), (6,2), (8,4) and (10,5)] are displayed.
In general, the spectra of each component
of the second-order optical susceptibility for all the chiral BN-NTs are similar,
except that the magnitude decreases somewhat as the diameter of the tubes increases.
The decrease of the magnitude with the tube diameter is particularly apparent for the
$xyz$ and $zzz$ components. 
For all the chiral BN-NTs, remarkably, the $zzz$ component is nearly ten times
larger than all the other nonvanishing components.
Another common feature is that both the real
and imaginary parts of $\chi^{(2)}(-2\omega,\omega,\omega)$ show a
rather oscillatory behavior, particularly for the $xyz$ and $xzx$ components
(Fig. 4). The amplitude of these oscillatory real and imaginary
parts is rather large in the photon energies of 2.0$\sim$ 6.0 eV
for the chiral BN-NTs with a small diameter. It should also be noted that 
the shape and magnitude of the $xzx$ and $zxx$ components look very much alike (Fig. 4b-c).

\begin{figure}[tb]
\includegraphics[width=8cm]{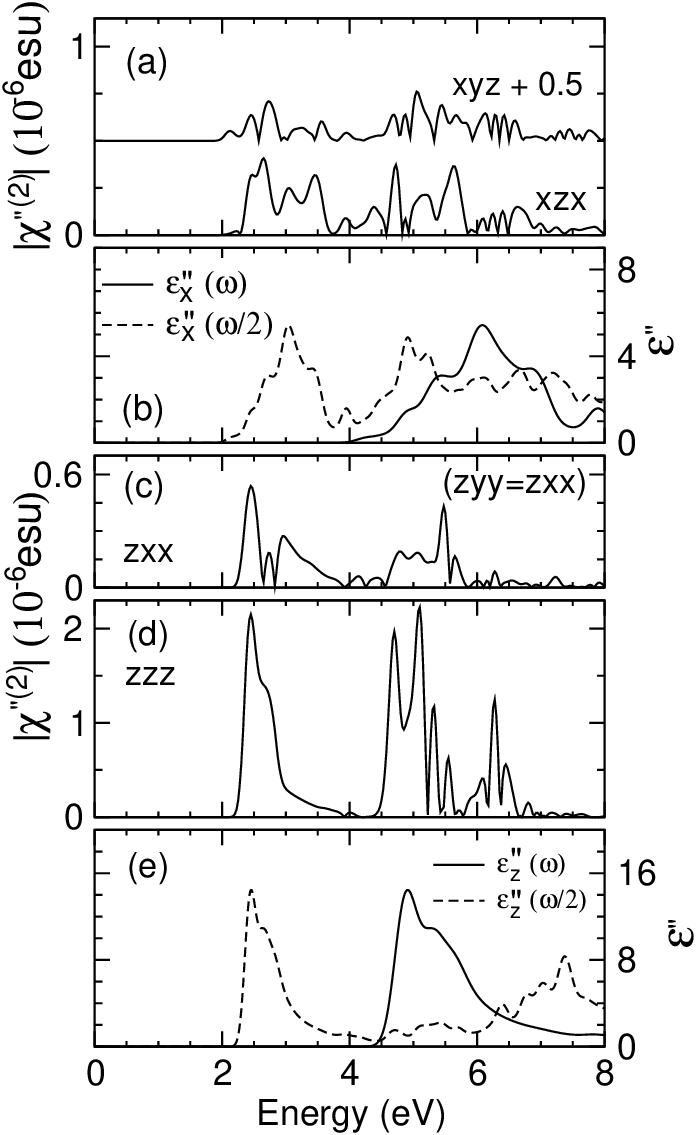}
\caption{\label{fig5} Absolute value of the imaginary part
of $\chi^{(2)}(-2\omega,\omega,\omega)$ (a, c, d) and 
$\varepsilon_a''(\omega)$ and $\varepsilon_a''(\omega/2)$ (imaginary part
of the dielectric function) (b, e) from Ref. ~\cite{guo05} of the zigzag (6,2)
BN nanotube.}
\end{figure}

Again, in order to understand the structures in the
calculated $\chi^{(2)}$ spectra of the chiral BN-NTs, the absolute values of the imaginary part
$\chi''^{(2)}$ of all the nonzero components of the (6,2) BN-NT are plotted and compared with the
absorptive part of the corresponding dielectric function
$\varepsilon''$ from our previous publication~\cite{guo05} in Fig. 5.
Remarkably, the first prominent peak between 2.0 and 4.0 eV
in the $\chi_{zzz}''^{(2)}$ spectrum is almost identical to the first peak in
the $\varepsilon_z''(\omega/2)$ (see Fig. 5d-e), indicating that it
is due to two-photon resonances. On the other hand, the second structure between 4.5 and 5.5 eV
in the $\chi_{zzz}''^{(2)}$ spectrum may be correlated with the
first peak in the $\varepsilon_z''(\omega)$, suggesting that
it is caused by the single-photon resonances. As for the zigzag BN-NTs,
both these single-photon and double-photon resonances
involve only interband $\pi \rightarrow \pi^{\ast}$
and $\sigma \rightarrow \sigma^{\ast}$ optical transitions for the
electric field vector ${\bf E}$ polarized parallel to the tube axis
($E \parallel \hat{z}$)~\cite{guo05}.
Fig. 5 further suggests that the feature between 2.0 and 4.0 eV in
the spectra of both the $\chi_{xzx}''^{(2)}$ and $\chi_{zxx}''^{(2)}$ as well 
as $\chi_{xyz}''^{(2)}$ may be attribued to the
two-photon resonances with $E\perp \hat{z}$ [cf. $\varepsilon_x''(\omega/2)$]
(see Fig. 5a-c), while, in contrast, the second feature in the photon energies above
4.5 eV is mainly due to the single-photon resonances
for $E\perp \hat{z}$ [cf. $\varepsilon_x''(\omega)$]. 
One exception is that the narrow peak at 2.5 eV in the $\chi_{zxx}''^{(2)}$ spectrum
(Fig. 5c) which may come predominantly from the 
two-photon resonances with $E\parallel \hat{z}$ [cf. $\varepsilon_z''(\omega/2)$ in Fig. 5e].

\subsection{Linear electro-optical coefficient}

\begin{table}
\caption{Calculated static refraction index $n$, second-order optical
susceptibility $\chi^{(2)}$ and linear electro-optical 
coefficient $r_{abc}$ of the zigzag and chiral BN nanotubes.
}
\begin{ruledtabular}

\begin{tabular}{c c c c}
      & $n_{x}$ ($n_{z}$) 
      & $\chi^{(2)}_{xzx}$, $\chi^{(2)}_{zxx}$, $\chi^{(2)}_{zzz}$ 
      & $r_{xzx}$, $r_{zxx}$, $r_{zzz}$ \\ 
      &     &  (pm/V) & (pm/V) \\ \hline
 (5,0)  & 1.88 (2.22) &  0.0,  0.0,   0.0 & 0.0, -0.0, 0.0 \\
 (6,0)  & 1.89 (2.21) & 29.1, 33.2, -117.9 & -3.34, -3.81, 9.88 \\
 (8,0)  & 1.90 (2.20) & 32.1, 32.3, -97.6 & -3.67, -3.70, 8.33 \\
 (9,0)  & 1.89 (2.17) &  0.0,  0.0, 0.0 &  0.0,  0.0, 0.0 \\
 (12,0)  & 1.89 (2.17) & 33.5, 33.2, -81.2 & -3.98, -3.95, 7.32 \\
 (13,0)  & 1.91 (2.18) & 35.3, 35.1, -83.0 & -4.07, -4.05, 7.35 \\
 (15,0)  & 1.89 (2.17) & 34.9, 34.6, -78.8 & -4.15, -4.11, 7.11 \\
 (16,0)  & 1.91 (2.18) & 36.4, 36.5, -81.1 & -4.20, -4.21, 7.18 \\
 (17,0)  & 1.91 (2.18) & 36.1, 36.3, -79.8 & -4.16, -4.19, 7.06 \\
 (20,0)  & 1.87 (2.16) & 30.0, 27.7, -61.8 & -3.68, -3.40, 5.58 \\
 (21,0)  & 1.90 (2.18) & 35.7, 35.8, -76.5 & -4.16, -4.17, 6.77 \\
 (24,0)  & 1.90 (2.18) & 35.8, 35.6, -75.2 & -4.17, -4.15, 6.66 \\
 (25,0)  & 1.90 (2.17) & 33.7, 33.8, -73.0 & -3.96, -3.98, 6.58 \\ 
 (27,0)  & 1.89 (2.16) &  0.0,  0.0, 0.0 &  0.0,  0.0, 0.0 \\  \hline
 (12,0)@(20,0) & 1.95 (2.20) & 10.1, 10.3, -19.3 &  -1.10, -1.12, 1.65 \\ 
 (12,0) bundle & 1.92 (2.19) & 37.0, 38.3, -89.0 &  -4.19, -4.33, 7.74 \\ \hline
 (4,2)  & 1.92 (2.14)  & 11.5, 11.5, -61.8 & -1.36, -1.36, 5.89\\
 (6,2)  & 1.89 (2.15)  & 15.8, 15.8, -64.9 & -1.91, -1.91, 6.07\\
 (8,4)  & 1.89 (2.15)  & 16.0, 16.0, -40.6 & -1.94, -1.94, 3.80\\
 (10,5)  & 1.89 (2.15) & 17.5, 17.5, -41.3 & -2.12, -2.12, 3.87\\
\end{tabular}
\end{ruledtabular}
\end{table}

In Table II, the calculated zero frequency linear electro-optic coefficient
$r(0)$ as well as the corresponding second-order nonlinear optical susceptibility
$\chi^{(2)}(0,0,0)$ are listed. The $r(0)$ is calculated from the 
corresponding $\chi^{(2)}(0,0,0)$ by using Equ. 7. 
The refraction index $n(0) (=\sqrt{\varepsilon(0)})$
is derived from the calculated static dielectric constant
$\varepsilon(0)$ which has been reported in our recent publication~\cite{guo05}.
Table II shows that apart from the (5,0), (9,0) and (27,0) BN-NTs which
have no nonvanishing $r_{abc}(0)$, all the other zigzag BN-NTs have very similar
linear electro-optical coefficients, as for the static dielectric constant
and polarizability~\cite{guo05}. Nevertheless, $r_{zzz}(0)$ decreases
slightly as the diameter increases, while $r_{xzx}(0)$ and $r_{zxx}(0)$ increase
slightly as the diameter goes up. We note that $\chi_{xzx}^{(2)}$ and $\chi_{zxx}^{(2)}$
are virtually identical, while $\chi_{zzz}^{(2)}$ is about two times larger
than $\chi_{xzx}^{(2)}$ and $\chi_{zxx}^{(2)}$. 

The chiral BN-NTs have two additional nonvanishing components $\chi_{xyz}^{(2)}$
and $\chi_{yzx}^{(2)}$. Nevertheless, the calculated static
values of $\chi^{(2)}_{xyz}$ and $\chi^{(2)}_{yzx}$ are zero, satisfying
the requirement by the so-called Kleinman symmetry~\cite{kle62} which
demands that $\chi^{(2)}_{xyz}(0)  = \chi^{(2)}_{yzx}(0)$. Consequently, the corresponding static
linear electro-optical coefficients $r_{xyz}(0)$ and $r_{yzx}(0)$ are zero too.
Therefore, $\chi_{yzx}^{(2)}$, $\chi^{(2)}_{xyz}$, $r_{xyz}(0)$ and $r_{yzx}(0)$ 
for the chiral BN-NTs are not listed in Table II.
As for the nonvanishing static components, $r_{xzx}(0)$ and $r_{zxx}(0)$ for the
(6,2), (8,4) and (10,5) BN-NTs are rather close, while $r_{zzz}(0)$ for the (6,2) BN-NT is
nearly 1.6 times larger than that for the (8,4) and (10,5) BN-NTs. Though $r_{zzz}(0)$ 
for the (4,2) BN-NT is close to that of the (6,2) BN-NT, $r_{xzx}(0)$ and $r_{zxx}(0)$ for the
(4,2) is somewhat smaller than that of the other armchair BN-NTs. 
 
Group III nitrides are promising materials for nonlinear optical
and opto-electric applications because of their unique physical
properties such as a wide band gap. However, the static $\chi^{(2)}(0)$
and hence $r(0)$  of bulk BN structures are the smallest among the group III
nitrides (see, e.g., ~\cite{gav00}). It is therefore remarkable that
BN in the single-walled zigzag BN-NTs and their bundles has greatly
enhanced static second-order nonlinear optical susceptibility
and linear electro-optical coefficient. This enhancement can be 
as large as thirty folds (see Table II and Ref. ~\cite{gav00}),
and it could make the BN nanotube structures have the largest 
$\chi^{(2)}$ and $r$ among the nitrides. In fact, the static $\chi^{(2)}(0)$
of the zigzag BN-NTs is up to five times larger than that of 
GaN which has the largest $\chi^{(2)}(0)$ among the bulk group III nitrides.~\cite{gav00}

\subsection{Effects of interwall interactions}

We have so far focused only on the nonlinear optical properties of single-walled BN-NTs.
However, BN-NTs are usually multiwalled or in the format of bundles. 
Therefore, to study the possible 
effects of interwall interactions on the nonlinear optical properties of BN-NTs,
we have also considered a double-walled BN-NT, namely, the zigzag (12,0)@(20,0),
and a single-walled zigzag (12,0) BN-NT bundle.
The double-walled nanotube is chosen because BN-NTs typically have a zigzag
structure, and also the interwall distance of about 3.2 \AA
$ $ between the (12,0) and (20,0) nanotubes is close to the interlayer distance 
in $h$-BN. The BN-NT bundle was modelled by a two-dimensional hexagonal array with the initial
interwall distance between adjacent nanotubes being about 3.2 \AA. 
The atomic positions and lattice constants were 
then fully relaxed by a conjugate gradient technique. 
Theoretical equilibrium nanotube structures were obtained
when the forces acting on all the atoms and the uniaxial stress were less than 0.03 eV/\AA$ $
and 2.0 kBar, respectively. The theoretical determined atomic structures were used in the
electronic structure and optical property calculations.
The calculated density of states and dielectric function of the (12,0)@(20,0) double-walled BN-NT 
have already been reported in Ref. ~\cite{guo05}.

\begin{figure}[tb]
\includegraphics[width=8cm]{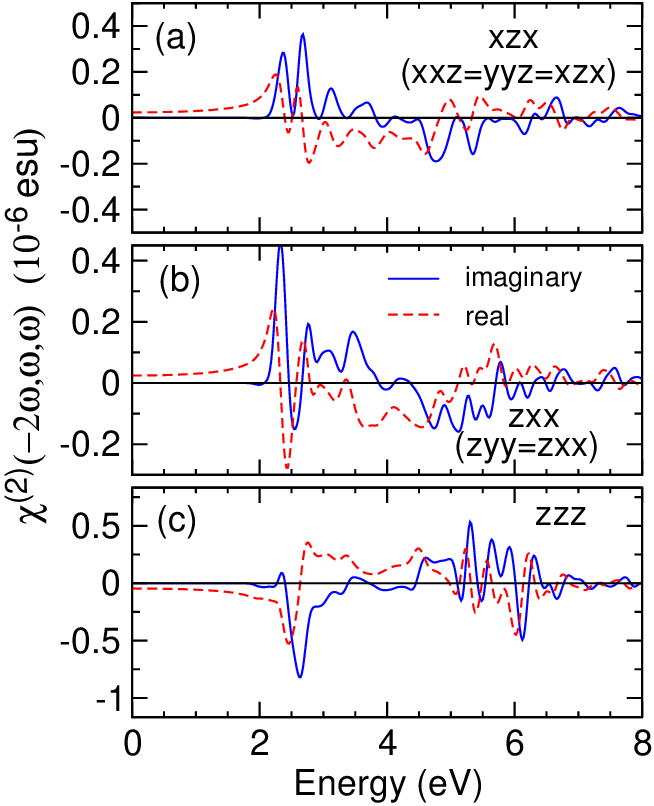}
\caption{\label{fig6} (Color online)  Real and imaginary parts
of $\chi^{(2)}(-2\omega,\omega,\omega)$ of the double-walled (12,0)@(20,0)
BN nanotube.}
\end{figure}

The calculated second-order optical susceptibility of the double-walled (12,0)@(20,0) BN-NT
is shown in Fig. 6. A uniform grid of $1\times 1\times 60$
in the Brillouin zone was used. Figs. 2 and 6 show that broadly speaking, each component
of the second-order optical susceptibility of the (12,0)@(20,0) BN-NT is rather similar
to the corresponding component of the (12,0) and (20,0) BN-NTs. However, the magnitude of
the second-order optical susceptibility of the double-walled BN-NT is significantly reduced in comparison
with that of the (12,0) and (20,0) BN-NTs. In particular, the magnitude of the
$zzz$ component of the (12,0)@(20,0) BN-NT is about three and two times smaller
than that of the (12,0) and (20,0) BN-NTs, respectively (Figs. 2 and 6).
In contrast, the density of states and dielectric function of the (12,0)@(20,0) as well as
(12,0) and (20,0) BN-NTs are very similar,~\cite{guo05} indicating that
the effects of interwall interaction
on the electronic structure and linear optical properties would be small.
In particular, the dielectric function
of the (12,0)@(20,0) is nearly the same as that of the (20,0).
The reduction in the magnitude of $\chi^{(2)}$ may be
explained by the fact that bulk $h$-BN does not show second-order nonlinear optical
behavior and that a multi-walled BN-NT with a very large diameter is essentially equivalent
to $h$-BN. Another discernable difference is that the second-order optical susceptibility 
of the (12,0)@(20,0) BN-NT is considerably more oscillatory than that of 
the (12,0) and (20,0) BN-NTs. 

The static SHG and linear electro-optic coefficients are compared with the single-walled
BN-NTs in Table II. Although the static refraction index of the (12,0)@(20,0) BN-NT
is almost identical to that of the single-walled BN-NTs, the static SHG coefficient
is more than three times smaller than  that of the single-walled zigzag nanotubes. Consequently,
the linear electro-optical coefficient of the double-walled nanotube is several times
smaller than that of the single-walled zigzag nanotubes (Table II).
 
\begin{figure}[tb]
\includegraphics[width=8cm]{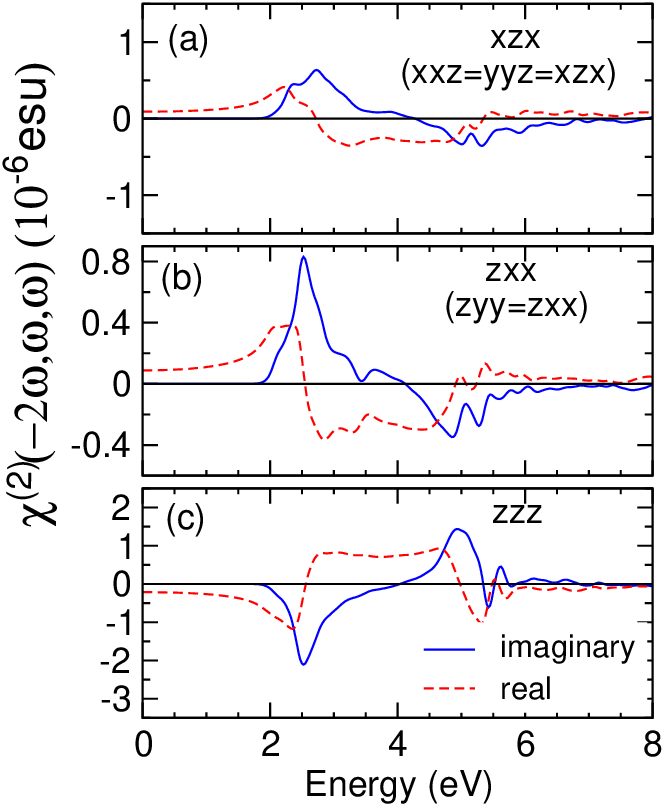}
\caption{\label{fig7} (Color online) Real and imaginary parts
of $\chi^{(2)}(-2\omega,\omega,\omega)$ of the single-walled (12,0)
BN-NT bundle.}
\end{figure}

The calculated second-order optical susceptibility of the (12,0) BN-NT bundle
is shown in Fig. 7. A uniform grid of $6\times 6\times 60$
in the Brillouin zone was used. The effective unit cell volume is used in order to 
compare quantitatively the optical properties of the bundle with that of
the isolated (12,0) BN nanotube. Nevertheless, the use of the solid
unit cell volume of the bundle would
only uniformly reduce the amplitude of the calculated second-order optical susceptibility 
by a factor of 1.43 but would not change its shape. Figs. 2 and 7 show that each component
of the second-order optical susceptibility of the (12,0) BN-NT bundle is similar
to the corresponding component of the isolated single-walled (12,0) BN-NT in
both the shape and magnitude. 
In particular, the $zzz$ component of the (12,0) BN-NT bundle is nearly identical to
that of the isolated (12,0) BN-NT (Figs. 2 and 7).
This clearly shows that the interwall interaction between the BN-NTs in the bundle
has essentially no effect on the second-order nonlinear optical properties, in strong
contrast to the case of the multi-walled BN-NTs.   
We also find that the dielectric function
of the (12,0) BN-NT bundle (not shown here) is nearly the same as that of the isolated
single-walled (20,0) BN-NT.
Nevertheless, there are minor differences in the second-order optical susceptibility between
the isolated (12,0) BN-NT and the (12,0) BN-NT bundle. For example, 
the second-order optical susceptibility
of the isolated (12,0) BN-NT is somewhat more oscillatory than that of
the (12,0) BN-NT bundle. This minor difference might be due to the fact that
more $k$-points for the (12,0) BN-NT bundle were used than for the single (12,0) BN-NT. 

The static SHG and linear electro-optic coefficients are compared with the single-walled
BN-NTs in Table II. Unlike the case of the double walled BN-NT, the static SHG linear and
electro-optic coefficients of the single-walled (12,0) BN-NT bundle
are rather similar to that of the single-walled BN-NTs (Table II).
This clearly shows that the interwall interaction hardly has any effect on 
the nonlinear optical properties, in strong contrast to the double walled BN-NT where
the interwall interaction significantly reduce the SHG and linear electro-optic coefficients.
Indeed, the SHG and linear electro-optic coefficients of the (12,0) BN-NT bundle are
slightly enhanced, as compared with that of the single (12,0) BN-NT (Table II). 
This is rather significant because in the nonlinear optical applications, 
one would usually need the BN-NTs in the format of either multiwalled BN-NTs or
BN-NT bundles. The present work therefore suggests that the single-walled BN-NT bundles
would be prefered for nonlinear optical and electro-optical applications.  

\section{Summary}

We have carried out a systematic {\it ab initio} study of
the second-order nonlinear optical properties of the BN-NTs
within density functional theory in the local density approximation.
We used the highly accurate full-potential PAW method.
The underlying atomic structure of the BN nanotubes was
determined theoretically.
Specifically, the properties of the single-walled zigzag 
[(5,0), (6,0), (8,0), (9,0), (12,0), (13,0), (15,0), (16,0), (17,0), 
(20,0), (24,0), (25,0), (27,0)], armchair [(3,3), (4,4),
(5,5), (6,6), (8,8), (12,12), (15,15)], and chiral [(4,2), (6,2), (8,4), (10,5)] 
nanotubes as well as the double-walled (12,0)@(20,0) nanotube and the 
single-walled (12,0) nanotube bundle have been calculated. 
For comparison, the second-order nonlinear optical properties 
of $h$-BN and the single BN sheet have also been 
calculated. Interestingly, we find that though $h$-BN has zero
second-order nonlinear optical susceptibility,
the $\chi^{(2)}_{aab}$, $\chi^{(2)}_{baa}$ and
$\chi^{(2)}_{bbb}$ for the isolated BN sheet are large and
generally several ten times larger than that of bulk BN in
both the  zinc-blende and wurtzite structures. 
We also find that, unlike carbon nanotubes, both the chiral and zigzag BN-NTs have
pronounced second-harmonic generation and linear electro-optical coefficients
which are comparable to that of the single BN sheet. The prominant structures in 
the spectra of $\chi^{(2)}(-2\omega,\omega,\omega)$
of the BN-NTs have been successfully correlated with the features in the corresponding
linear optical dielectric function $\varepsilon (\omega)$ in terms of
single-photon and double-photon resonances. 
Though the interwall interaction in the double-walled BN-NTs is found to reduce
the second-order nonlinear optical coefficients significantly, the interwall interaction
in the single-walled BN-NT bundle has essentially no effect on the nonlinear optical
properties, suggesting that the single-walled BN-NTs and single-walled BN-NT bundles 
are promising nonlinear optical materials for applications in, e.g., second-harmonic generation,
sum frequency generation and electro-optical switches. We hope that this work would stimulate
experimental investigations into the second-order nonlinear optical properties of BN-NTs.

\section{Acknowledgments}

The authors thank Ding-Sheng Wang and Chung-Gang Duan for helpful discussions
and for allowing to access their nonlinear optical calculation program.
The authors gratefully acknowledge financial supports from National Science Council,
Ministry of Economic Affairs (93-EC-17-A-08-S1-0006) and 
NCTS/TPE of the Republic of China. They also thank National Center for High-performance
Computing of Taiwan for providing CPU time.\\

\end{document}